\documentclass[12pt,twoside]{article}
\usepackage{latexsym,amsmath,amssymb,amsthm,amsfonts,graphicx}
\usepackage{natbib}
\bibpunct{(}{)}{;}{a}{,}{,}

\def\,{\mskip 3mu} \def\>{\mskip 4mu plus 2mu minus 4mu} \def\;{\mskip 5mu plus 5mu} \def\!{\mskip-3mu}
\def\dispmuskip{\thinmuskip= 3mu plus 0mu minus 2mu \medmuskip=  4mu plus 2mu minus 2mu \thickmuskip=5mu plus 5mu minus 2mu}
\def\textmuskip{\thinmuskip= 0mu                    \medmuskip=  1mu plus 1mu minus 1mu \thickmuskip=2mu plus 3mu minus 1mu}
\textmuskip
\def\beq{\dispmuskip\begin{equation}}    \def\eeq{\end{equation}\textmuskip}
\def\beqn{\dispmuskip\begin{displaymath}}\def\eeqn{\end{displaymath}\textmuskip}
\def\bqa{\dispmuskip\begin{eqnarray}}    \def\eqa{\end{eqnarray}\textmuskip}
\def\bqan{\dispmuskip\begin{eqnarray*}}  \def\eqan{\end{eqnarray*}\textmuskip}

\newtheorem{theorem}{Theorem}

\newtheorem{lemma}[theorem]{Lemma}

\newtheorem{principle}[theorem]{Principle}
\newenvironment{keywords}{\centerline{\bf\small
Keywords}\begin{quote}\small}{\par\end{quote}\vskip 1ex}

\newtheorem{myexample}[theorem]{Example}

\def\paradot#1{\vspace{1.3ex plus 0.7ex minus 0.5ex}\noindent{\bf\boldmath{#1.}}}

\def\fr#1#2{{\textstyle{#1\over#2}}}

\def\SetR{\mathbb{R}}

\def\P{{\rm P}}                         

\def\v{\boldsymbol}
\def\trp{{\!\top\!}}

\def\a{\alpha}

\def\s{\sigma}

\def\b{\beta}
\def\l{\lambda}

\def\M{{\cal M}}

\def\X{{\cal X}}
\def\Y{{\cal Y}}
\def\S{{\cal S}}

\def\L{\text{\rm Loss}}

\def\Rank{\text{\rm Rank}}
\def\LR{\text{\rm LR}}
\def\tr{\text{\rm tr}}

\def\KL{\text{\rm KL}}
\def\BIC{\text{\rm BIC}}

\def\GCV{\text{\rm GCV}}

\def\cov{\text{\rm cov}}

\def\argmin{\text{\rm argmin}}
\def\argmax{\text{\rm argmax}}

\def\df{\text{\rm df}}
\def\DF{\text{\rm DF}}
\def\diag{\text{\rm diag}}

\begin{document}

\title{\bf\Large\hrule height5pt \vskip 4mm
The Loss Rank Criterion for Variable Selection in Linear Regression Analysis
\vskip 4mm \hrule height2pt}

\author{\\
{\bf Minh Ngoc Tran}\footnote{The author would like to thank the editor, an associate editor and two anonymous reviewers for careful reading and constructive comments which helped improve the paper greatly. 
The author is also grateful to David J. Nott for helpful suggestions which led to a better presentation of the paper.}\\[3mm]
\normalsize Department of Statistics and Applied Probability\\
\normalsize National University of Singapore, Singapore 117546\\
\normalsize \texttt{ngoctm@nus.edu.sg} \\[1ex]
}%
\maketitle

\begin{abstract}\noindent
Lasso and other regularization procedures are attractive methods for variable selection,
subject to a proper choice of shrinkage parameter. 
Given a set of potential subsets produced by a regularization algorithm,
a consistent model selection criterion is proposed to select the best one among this preselected set.
The approach leads to a fast and efficient procedure for variable selection, especially in high-dimensional settings. 
Model selection consistency of the suggested criterion is proven when the number of covariates $d$ is fixed. 
Simulation studies suggest that the criterion still enjoys model selection consistency when $d$ is much larger than the sample size.
The simulations also show that our approach for variable selection works surprisingly well in comparison with existing competitors.
The method is also applied to a real data set. 
\end{abstract}

\begin{keywords}
Model selection,
lasso,
loss rank principle,
shrinkage parameter,
variable selection
\end{keywords}

\newpage
\section{Introduction}\label{intr}
Variable selection is probably the most fundamental and important topic in linear regression analysis.
We consider the case where a large number (even larger than the sample size) 
of candidate covariates are introduced at the initial stage of modeling.
One then has to select a smaller subset of the covariates to fit/interpret the data.
If the number of potential covariates is not so large (as small as 30), 
one may use subset selection to select significant variables \citep{Miller:90}.
However, with a large number of covariates,
searching on model space is computationally infeasible.
Lasso \citep{Tibshirani:96} and other regularization procedures (e.g., the adaptive lasso of \cite{Zou:06}, SCAD of \cite{Fan:01}) 
are successful methods to overcome this problem.
A Lasso-type procedure estimates the regression coefficient vector $\v\beta$ by minimizing
the sum of the squared error and a regularization term
\beq\label{PenalyCommonForm}
\|\v y-X\v\beta\|^2+\l T(\v\b),
\eeq
where $X$ is an $(n\times d)$ non-random design matrix,
$\v y$ is an $n-$vector of responses,
and $\l\geq0$ is a shrinkage parameter that controls the amount of regularization.
The regularization function $T(\v\b)$ can take different forms according to different regularization procedures.
The original and most popular one used in Lasso is the $l_1$ norm $T(\v\b)=\sum_{j=1}^d|\beta_j|$.
As $\l$ increases, the coefficients are continuously shrunk towards 0. 
When $\l$ is sufficiently large, some coefficients are shrunk to exact 0, thus leading to sparse solutions.
This feature makes the Lasso-type procedures very attractive for variable selection.
Indeed, their model selection consistency has been shown
\citep{Zhao:06,Meinshausen:06,Fan:01}:
Under some conditions, there exists a ``proper" sequence of shrinkage parameters $\{\l_n\}$
under which
\beq\label{consistency}
\{j:\hat\beta^{\l_n}_{j}\not=0\}=\S_T\;\;\text{w.p.1 when sample size $n$ is large enough},
\eeq
where $\hat{\v\beta}_{\l_n}=(\hat\beta^{\l_n}_{1},...,\hat\beta^{\l_n}_{d})^\trp$ 
is the regularized estimator of $\v\beta$ with shrinkage parameter $\l_n$,
and $\S_T$ is the true model, i.e., $\S_T$ is the index set of true covariates.
Therefore, it is convenient to use the Lasso-type procedures for variable selection purposes.

The remaining problem in practice is how to choose such proper $\l_n$. 
A widely-used criterion is the generalized cross-validation criterion (GCV) \citep{Craven:79,Tibshirani:96}.
However, theoretical properties of GCV for choosing $\l$ for the purpose of variable selection
have not been investigated yet.
Furthermore, for choosing shrinkage parameter for the SCAD method \citep{Fan:01},
a regularization method closely related to Lasso, 
GCV seems to be likely to choose shrinkage parameters that produce overfitted models \citep{Wang:07}.
\cite{Hui:07} showed that the number of nonzero coefficients is an unbiased estimate for the degrees
of freedom of the lasso.
As a result, popular model selection criteria - like AIC, BIC and $C_p$ - can be used for selecting $\l$.
However, theoretical properties of the selected model remain unknown. 
The main contribution of this paper is to propose a criterion 
for selecting shrinkage parameters
in order for regularization procedures to produce the true model.
(Throughout this paper, the true model is assumed to exist.
We note, however, that whether or not the true model exists is still a controversial issue
in the model selection literature, see \cite{Burnham:02}).

Although regularization procedures can be used for simultaneous variable selection and estimation,
it seems to be impossible to tune the shrinkage parameter to achieve both model selection consistency
and optimal estimation at the same time.
For an orthogonal design, \cite{Leng:06} showed that 
the Lasso estimator that is optimal in terms of estimation does not give consistent model selection.
This fact was also shown by \cite{Poetscher:09} for other regularized estimators.
We are in this paper primarily concerned with the problem of variable selection,
i.e., we use a Lasso-type procedure to produce a set of potential subsets
and then select the best one among this preselected set using a model selection criterion.
It was brought to our attention by a reviewer that the idea of using Lasso
as a ``selector" was also briefly mentioned in \cite{Friedman:08,Efron:04}.
They discussed an approach to reduce estimation bias on the non-zero estimated coefficients 
in which the Lasso (with some method for choosing the shrinkage parameter) is used as a subset selector 
and then a different unpenalized procedure is used to estimate the coefficients w.r.t. the selected covariates.
Our approach is to select the best subset of covariates - using 
a model selection criterion as the stopping rule - among a preselected set of potential subsets
produced by a Lasso-type procedure.
The preselected set consists of at most $d$ subsets rather than $2^d$ possible subsets if using subset selection.
After selecting the best subset, we of course can use an unpenalized procedure to estimate the coefficients
in order to reduce estimation bias. 

The model selection criterion we use is derived from the {\it loss rank principle} (LoRP),
a general-purpose principle for model selection,
introduced recently by \cite{Hutter:07,Hutter:09}.
LoRP selects a model that has the smallest {\it loss rank}.
The loss rank of a model is defined as the number of other ``fictitious" data
that fit the model better than the training data (see Section 2 for a formal introduction).
It was shown by \cite{Hutter:09} that minimizing the
loss rank is a suitable criterion for model selection, 
since it trades off between the quality of fit and the model complexity.
LoRP seems to be a promising principle with enormous potential, leading to a rich field.
\cite{Tran:09a} demonstrated the use of LoRP for selecting the ridge parameter in ridge regression.
\cite{Tran:09b} adapted the idea of LoRP for model selection in a classification context 
and showed its close connection with excellent model selection techniques based on Rademacher complexities \citep{Koltchinskii:01,Bartlett:02}.
In this paper, we shall show that LoRP also successfully applies to selecting the shrinkage parameter
for the purpose of variable selection.

The main contribution of this paper is to propose a criterion, called the loss rank (LR) criterion, 
for selecting shrinkage parameters for variable selection purposes.
As long as the regularization procedure in use has the consistency property \eqref{consistency},
the shrinkage parameter selected by the LR criterion will produce the true model asymptotically with probability 1.
This model selection consistency of the proposed criterion will be proven theoretically
in the case where the number of covariates $d$ is fixed and smaller than $n$.
For cases with $d\gg n$, our simulation study suggests that this property still holds.
The simulation also shows that our method for variable selection works surprisingly well.
Benefiting from fast $l_1$-regularization algorithms,
our method is able to correctly identify significant variables from thousands of candidates in several CPU seconds. 
 
The paper is organized as follows. 
The main idea of LoRP is briefly reviewed in Section \ref{secLoRP}.
The LR criterion is derived and its model selection consistency is proven in Section \ref{secLRcriterion}.
Simulation studies and real-data application are presented in Section \ref{Secsimulation}.
Section \ref{secCon} contains the conclusions and outlook.
The proofs are relegated to the appendix.

\section{The loss rank principle}\label{secLoRP}
In this section, we give a brief review of the loss rank principle (LoRP).
The reader is referred to \cite{Hutter:07,Hutter:09} for the details.

Let us consider a training data set 
$D=(\v x,\v y)=\{(x_1,y_1),...,(x_n,y_n)\}\in(\X\times \Y)^n$ from a regression model
\beqn
y_i=f(x_i)+\epsilon_i.
\eeqn
We first consider discrete $\Y$. Suppose that we use a model $M$ to fit the data $D$, 
e.g., $M$ is a linear regression model with $d$ covariates, 
or $M$ is a $k$-nearest neighbors regression model. 
Imagine that in experiment situations we can conduct the experiment many times 
with fixed design points $\v x$. 
We then would get many other (fictitious) output $\v y'$. 
Observe that if the model $M$ is complex/flexible (large $d$, small $k$), 
then $M$ fits the training data $(\v x,\v y)$ well and it also fits $(\v x,\v y')$ well
(with respect to some loss function).
Here, for simplicity, we only consider the squared loss 
$\L_M(\v y|\v x)=\|\v y-\hat{\v y}\|^2=\sum_1^n(y_i-\hat{y}_i)^2$ where $\hat{\v y}$ 
is the fitted vector under model $M$. 
Therefore the loss rank of $M$ defined by
\beqn
\Rank_M(D):=\#\{\v y'\in\Y^n:\L_M(\v y'|\v x)\leq \L_M(\v y|\v x)\}
\eeqn    
will be large for complex $M$. 
Conversely, as argued by \cite{Hutter:09}, if $M$ is small/rigid, 
that both $\L_M(\v y|\v x)$ and $\L_M(\v y'|\v x)$ are large also leads 
to a large loss rank. Thus, it is natural to choose a model 
with the smallest loss rank as a good model.
By doing this, we trade off between the fit and the model complexity.

In the case of continuous $\Y$, say for instance $\Y=\SetR$, 
it is natural to use the concept of volume instead of the counting measure 
in the definition of loss rank, i.e.,
\beqn
\Rank_M(D):=\text{Vol}\{\v y'\in\SetR^n:\L_M(\v y'|\v x)\leq \L_M(\v y|\v x)\}.
\eeqn    
Consider the case of linear models where the fitted vector 
$\hat{\v y}$ is linear in $\v y$, i.e., $\hat{\v y}= M(\v x)\v y$ 
where the regression matrix $ M= M(\v x)$ depends only on $\v x$
(using the same symbol $M$ for both model and regression matrix will not cause any confusion in the following).
Then $\L_M(\v y|\v x)=\|\v y- M\v y\|^2=\v y^\trp A\v y$ with 
$ A=(I_n- M)^\trp(I_n- M)$ and the loss rank is
\beqn
\Rank_M(D)=\text{Vol}\{\v y'\in\SetR^n:\v y'^\trp A\v y'\leq L\}\ \text{where}\ L:=\v y^\trp A\v y.
\eeqn    
Suppose at the moment that $\det( A)\not=0$.
The set $\{\v y'\in\SetR^n:\v y'^\trp A\v y'\leq L\}$ 
is an ellipsoid in $\SetR^n$, so that its volume is
\beqn
\Rank_M(D)=\dfrac{v_nL^{n/2}}{\sqrt{\det A}}
\eeqn
where $v_n=\pi^{n/2}/\Gamma(\fr n2+1)$ is the volume of the unit sphere in $\SetR^n$. 
Because the logarithm is monotone increasing and $v_n$ depends only on $n$, 
it is equivalent to consider
\beq\label{equ5}
\LR_M(D)=\fr n2\log(\v y^\trp A\v y)-\fr 12\log\det A.
\eeq 
\begin{principle}\label{principle1}
Given a class of linear models $\mathcal M=\{M\}$,
the best model among $\M$ is the one with the smallest loss rank
\beq\label{equ6}
M^{best}=\argmin_{M\in\M}\{\fr n2\log(\v y^\trp A\v y)-\fr 12\log\det A\}
\eeq
where $ A=(I_n- M)^\trp(I_n- M)$, provided that $\det A>0$.
\end{principle}
Now we consider the case where $\det A=0$ (e.g., projective regression) or 
$A$ is nearly singular (e.g., ridge regression when the ridge parameter is very close to 0). 
In such cases, $\Rank_M(D)$ is infinity or extremely large.
Following the principle of ridge regression, we add, in order to prevent the loss rank
from being infinity or extremely large, a small penalty $\a\|\v y\|^2$ to the loss
\beqn
\L_M^\a(\v y|\v x):=\|\hat{\v y}-\v y\|^2+\a\|\v y\|^2=\v y^\trp S_\a\v y,\;\;  S_\a= A+\a I_n
\eeqn 
where $\a>0$ is a small number to be determined later. 
Now, $S_\a$ being not singular yields
\beqn
\Rank_M^\a(D)=\text{Vol}\{\v y'\in\SetR^n:\v y'^\trp S_\a\v y'\leq L\}
=\dfrac{v_nL^{n/2}}{\sqrt{\det S_\a}}\ \text{where}\ L:=\v y^\trp S_\a\v y.
\eeqn
Taking logarithm and neglecting a constant independent of $M$, 
we define the loss rank of model $M$ (dependent on $\a$) as
\beq\label{equ7}
\LR_M^\a(D)=\fr n2\log(\v y^\trp S_\a\v y)-\fr 12\log\det S_\a.
\eeq
How do we deal with the extra parameter $\a$? 
We are seeking a model of smallest loss rank, 
so it is natural to minimize $\LR_M^\a(D)$ in $\a$ 
(see \cite{Hutter:09} for a more detailed interpretation). 
Therefore, we finally define the loss rank of model $M$ as
\beq\label{equ8}
\LR_M(D)=\inf_{\a>0}\LR_M^\a(D)=\inf_{\a>0}\{\fr n2\log(\v y^\trp S_\a\v y)-\fr 12\log\det S_\a\}.
\eeq 
\begin{principle}(Hutter and Tran, 2010)\label{principle2} 
Given a class of linear models $\mathcal M=\{M\}$, 
the best model among $\M$ is the one with the smallest loss rank
\beq\label{equ9}
M^{best}=\argmin_{M\in\M}\LR_M(D)=\argmin_{M\in\M}\inf_{\a>0}\{\fr n2\log(\v y^\trp S_\a\v y)-\fr 12\log\det S_\a\}
\eeq
where $ S_\a= A+\a I_n=(I_n- M)^\trp(I_n- M)+\a I_n$.
\end{principle}
Many attractive properties of LoRP have been pointed out: 
LoRP reduces to Bayesian model selection in some special cases and
has an interpretation in terms of MDL principle \citep{Hutter:09};
in the classification context, 
LoRP has a close connection with (and works in some cases better than) model selection techniques
based on Rademacher complexities \citep{Tran:09b}.
By virtue of LoRP, the loss rank criterion for 
selecting the Lasso parameter will be derived in the next section.

\section{The LR criterion}\label{secLRcriterion}
Let us go back to the variable selection problem in linear regression analysis.
We build on the notation of \cite{Hutter:09}.
Given a (large) set of $d$ potential covariates $X_1,...,X_d$ and a response variable $Y$,
we consider the problem of choosing the important covariates for explaining $Y$ as a linear function of $X_1,...,X_d$.
It is assumed as usual that the covariates are linearly independent
and that $E(Y|X_1,...,X_d)$ is a linear combination of $X_1,...,X_d$ with some of the coefficients are zero.
We are primarily interested in identifying the non-zero coefficients.

Suppose that the response vector $\v y$ and the design matrix $X$ have been centered, 
so that the intercept is omitted from models.
Denote by $\S_T=\{j_1^*,...j_{d^*}^*\}$ and $\S=\{j_1,...j_{d_0}\}$ the true model and a candidate model, respectively.
Under model $\S$, we can write
\beqn
\v y=X\v\beta_\S+\sigma\v\epsilon
\eeqn
where $E(\v\epsilon)=0,\ \cov(\v\epsilon)=I_n,\ \sigma>0$.
We shall consider 
$\v\beta_\S=(\beta_{\S 1},...,\beta_{\S d})^\trp$ as a point in $\SetR^d$ with $\beta_{\S j}=0$ if $j\not\in\S$.
Denote by $\Theta(\S):=\{\theta_\S=(\v\beta_\S,\sigma^2)\in\SetR^d\times\SetR_+\}$ the parameter space of model $\S$ and 
by $X_\S$ the $(n\times d_0)$ design matrix obtained from $X$ by
removing the $j$th column for all $j\not\in\S$.

\subsection{The LR criterion}
Let $\hat{\v\beta}_\l=(\hat\beta^{\l}_{1},...,\hat\beta^{\l}_{d})^\trp$ be the regularized estimator of 
$\v\beta$ w.r.t. a certain shrinkage parameter $\l$,
i.e., $\hat{\v\beta}_\l$ is the solution of \eqref{PenalyCommonForm}.
Denote by $\S_\l=\{j:\hat\beta^{\l}_j\not=0\}$
the index set corresponding to the non-zero coefficients, 
by $\df_\l=|\S_\l|$ the number of non-zero coefficients,
and by $X_{\S_\l}$ the design matrix corresponding to the selected covariates.
We assume at the moment that $\df_\l\leq n$ and further assume that matrices $X_{\S_\l}$ are full rank.
The case where $\df_\l>n$ will be dealt with later on.

Fitting model $\S_\l$ by least squares, we denote the OLS estimator
and the variance estimator by
\beqn
\hat{\v\beta}_{\S_\l}=(X_{\S_\l}^\trp X_{\S_\l})^{-1} X_{\S_\l}^\trp\v y\;;\; \hat{\sigma}_{\S_\l}^2=\fr1n\|\v y- X_{\S_\l}\hat{\v\beta}_{\S_\l}\|^2,
\eeqn
respectively. The fitted vector under model $\S_\l$
\beqn
\hat {\v y}_{\S_\l}= X_{\S_\l}\hat{\v\beta}_{\S_\l}= M_{\S_\l}\v y\;\text{with}\; M_{\S_\l}:= X_{\S_\l}( X_{\S_\l}^\trp X_{\S_\l})^{-1} X_{\S_\l}^\trp
\eeqn
is, conditionally on $\S_\l$, linear\footnote{Strictly speaking, $\hat {\v y}_{\S_\l}$ is not linear in $\v y$ because $\S_\l$ depends on $\v y$.
However, we can consider preselected subsets $\S_\l$ as fixed models.
If instead we first derive the LR criterion for a general fixed model $\S$ and then apply to $\S_\l$,
we get the same results.} in $\v y$. 
Then from \eqref{equ7}, the loss rank of model $\S_\l$ with 
parameter $\a$ is
\beqn
\LR_\l^\a\equiv\LR_{\S_\l}^\a=\fr n2\log(\v y^\trp S_\a^\l\v y)-\fr 12\log\det( S_\a^\l)
\eeqn
where $S_\a^\l=(I- M_{\S_\l})^\trp(I- M_{\S_\l})+\a I=(1+\a)I- M_{\S_\l}$.
Because projection matrix $M_{\S_\l}$ has $\df_\l$ eigenvalues $1$ and $n-\df_\l$ eigenvalues $0$,
$S_\a^\l$ has $\df_\l$ eigenvalues $\a$ and $n-\df_\l$ eigenvalues $1+\a$. Thus, $\det S_\a^\l=\a^{\df_\l}(1+\a)^{n-\df_\l}$.
Let $\rho_\l:=\|\v y-\hat {\v y}_{\S_\l}\|^2/\|\v y\|^2$, we have 
\beqn
\LR_\l^\a=\fr n2\log\v y^\trp\v y+\fr n2\log(\rho_\l+\a)-\fr{\df_\l}{2}\log\a-\fr{n-\df_\l}{2}\log(1+\a).
\eeqn
Taking derivative w.r.t $\a$, it is easy to see that 
$\LR_\l^\a$ is minimized at $\a_m=\fr{\rho_\l \df_\l}{(1-\rho_\l)n-\df_\l}$ provided that $1-\rho_\l>{\df_\l}/{n}$. 
This condition is ensured by Assumption (A3) below.
Finally, after some algebra, the loss rank of model $\S_\l$ as defined in \eqref{equ8} can be explicitly expressed as  
\beq\label{equLR}
\LR_\l=\LR_\l^{\a_m}=\fr n2\log\|\v y\|^2-\fr n2\KL(\fr{\df_\l}{n}\|1-\rho_\l).
\eeq
where $\KL(p\|q)=p\log\fr pq+(1-p)\log\fr{1-p}{1-q}$ is the Kullback-Leibler divergence between 
the Bernoulli distributions with parameters $p,q\in(0,1)$. 
The optimal shrinkage parameter(s) $\l$ (for variable selection purposes) chosen by the LR criterion will be 
\beq\label{LRC}
\hat\l_{\LR}\in\argmin_{\l\geq0}\LR_\l=\argmax_{\l\geq0}\KL(\fr{\df_\l}{n}\|1-\rho_\l).
\eeq
Often, $\LR_\l$ reaches its minimum in an interval $(\hat{\l}_{l},\hat{\l}_{u})$ (see Figure \ref{figure2}).
Any $\l$ in this interval produces the same model.
This can be explained as follows.
When $\l$ increases from 0 to infinity, the number of non-zero coefficients of $\hat{\v\beta}_\l$
will be a non-increasing step function of $\l$ \citep{Efron:04};
in other words, the covariates are in turn removed from the models.
As a result, by its definition $\LR_\l$ is also a step function. 
Note that our emphasis is on variable selection rather than on coefficient estimation.

\subsection{Model selection consistency for fixed $d$}
In order to prove the model selection consistency of the LR criterion,
we assume in this section that $d$ is fixed and $d\leq n$.
We need the following assumptions
\begin{itemize}
\item[(A1)] There exists a deterministic sequence of reference shrinkage parameters $\l_n$ such that $\S_{\l_n}\to\S_T$ w.p.1.
\item[(A2)] $\v\epsilon$ is Gaussian $N(0,I_n)$.
\item[(A3)] For each candidate $\l$, $\rho_\l$ is bounded away from 0 and 1, i.e.,
there are constants $c_1,\ c_2$ such that $0<c_1\leq\rho_\l\leq c_2<1$ w.p.1.
\end{itemize}

\paradot{Comments}
$\rho_\l=\|\v y-\hat {\v y}_{\S_\l}\|^2/\|\v y\|^2$ is a measure of fit.
In extreme cases where the resulting model $\S_\l$ is too big or too small,
$\rho_\l$ will be close to 0 and 1, respectively. 
Therefore, it is reasonable to consider only $\l$ in which $\rho_\l$ is bounded away from 0 and 1.
Note that for every $\S_\l$ we have that
\beqn
\rho_\l=\frac{\|\v y-\hat {\v y}_{\S_\l}\|^2}{\|\v y-\hat {\v y}_{\S_\l}\|^2+\|\hat {\v y}_{\S_\l}\|^2}=\frac{\hat\sigma^2_{\S_\l}}{\hat\sigma^2_{\S_\l}+\|\hat {\v y}_{\S_\l}\|^2/n}.
\eeqn
For $\l$ such that $\S_\l$ is the true model $\S_T$, (A3) follows from a mild sufficient condition
\beqn
0<\lim\inf_{n\to\infty}(\fr 1n\|\hat{\v y}_{\S_T}\|^2)\leq \lim\sup_{n\to\infty}(\fr 1n\|\hat{\v y}_{\S_T}\|^2)<\infty\;\;\text{and}\;\;\hat\sigma^2_{\S_T}\to\sigma^2>0\;\;\text{w.p.1}
\eeqn
where $\hat{\v y}_{\S_T}$ is the fitted vector under the true model.
Moreover, if the intercept is included in the models, we have that $n(\bar{\v y})^2\leq\|\hat{\v y}_{\S_\l}\|^2\leq\|\v y\|^2$.
(A3) then follows from a very mild condition
\beqn
0<\lim\inf_{n\to\infty}(\bar{\v y})^2\leq\lim\sup_{n\to\infty}(\fr1n\|\v y\|^2)<\infty\;\;\text{and}\;\;\hat\sigma_\S^2\to\text{constant}>0\;\forall\S\;\text{w.p.1}. 
\eeqn 
Assumption (A1) is satisfied by some regularization procedures, for example, Lasso \citep{Zhao:06} and SCAD \citep{Fan:01}.
Normality assumption (A2) is not a necessary condition for consistency.
This assumption can be relaxed, but then a more complicated proof technique is needed. 

We have the following lemma.

\begin{lemma}\label{lemma}
The loss rank of model $\S_\l$ can be rewritten as 
\beq\label{varsel}
  \LR_\l \;=\; \fr n2\log(n\hat\s^2_{\S_\l})+\fr n2 H(\fr{\df_\l}{n}) +\fr {\df_\l}2\log\fr{1-\rho_\l}{\rho_\l}
\eeq
where $H(p):=-p\log p-(1-p)\log(1-p)$ is the entropy of $p$. Under
Assumption (A3), 
the loss rank $\LR_\l$ has the form
\beq\label{equlemma}
  \LR_\l \;=\; \fr n2\log\hat\s_{\S_\l}^2+\fr{\df_\l}{2} \log n +\fr{n}{2}\log n+O_\P(1),
\eeq
where $O_\P(1)$ denotes a bounded random variable w.p.1.
\end{lemma}
\begin{proof}
With $\rho_\l=n\hat\sigma_{\S_\l}^2/\|\v y\|^2$, rearranging terms in \eqref{equLR} we get \eqref{varsel}.
The fact that $\fr {H(p)}{p}+\log p\to1$ as $p\to0$ implies that (note that $\df_\l\leq d$ and $d$ is fixed) 
\beqn
\fr n2 H(\fr{\df_\l}{n})=\fr{\df_\l}{2}\log n+\fr{\df_\l}{2}(1-\log\df_\l)+o(1).
\eeqn
Under Assumption (A3), the last term of \eqref{varsel} is bounded.
This completes the proof.
\end{proof}

The above lemma is used to prove model selection consistency of the LR criterion.
\begin{theorem}[Model selection consistency of the LR criterion]\label{LRconsistency}
Assume that $d$ is fixed.
Under Assumptions (A1)-(A3),
the shrinkage parameter selected by the LR criterion will produce the true model w.p.1 when $n$ is large enough, i.e.,
\beqn
\P(\S_{\hat\l_\LR}=\S_T)\to1
\eeqn
where $\hat\l_\LR$ is determined in \eqref{LRC}.
\end{theorem}
The idea of the proof is to bound the probabilities of
picking under- and overfitted models.
A model $\S$ is said to be underfitted if $\S$ misses at least one true covariate (i.e., $\S\not\supseteq \S_T$),
overfitted if $\S$ contains all true covariates and at least one untrue (i.e., $\S\supsetneq \S_T$).
There is a finite number of such $\S$, so it is sufficient to prove that $\P(\S_{\hat\l_\LR}=\S)\to0$ for each of them.
The detailed proof is relegated to the appendix.

We can of course use other model selection criteria rather than LoRP
for choosing the best subset among the preselected set produced by the regularization procedure.
The most widely-used selection criteria in statistics are probably AIC \citep{Akaike:73} and BIC \citep{Schwarz:78}.
AIC is asymptotically optimal in terms of loss efficiency but likely to select overfitted models,
while BIC is asymptotically optimal in terms of model selection consistency; see \cite{Shao:97,Yang:05}.  
Therefore one may use BIC as another stopping rule besides LoRP.
The shrinkage parameter chosen by BIC will be
\beq\label{BIC}
\hat\l_\BIC\in\argmin_{\l\geq0}\BIC_\l\;\;\text{where}\;\;\BIC_\l:=\fr n2\log\hat\s_{\S_\l}^2+\fr{\df_\l}{2} \log n.
\eeq  
We see from Lemma \ref{lemma} that, up to a constant, the LR criterion is asymptotically equivalent to BIC.
It follows from the proof of Theorem \ref{LRconsistency} that using BIC also leads to the same model selection consistency,
i.e., $\P(\S_{\hat\l_\BIC}=\S_T)\to1$ as $n\to\infty$.
However, finite-sample simulation studies in the next section show that the LR criterion works better than BIC,
especially when $d\gg n$. 

\subsection{Large $d$ small $n$}
High-dimensional variable selection problems in which $d\gg n$ is currently of great interest to scientists.
In order for such a problem to be solvable, an essential assumption needed is that it is $d^*-$sparse \citep{Candes:07},
i.e., the number of true covariates $d^*$ must be smaller than $n$.
Under this solvability assumption, it is clear that  
we can safely ignore irrelevant cases in which the number of covariates $\df_\l$ under consideration is larger than $n$.
Then the LR criterion \eqref{equLR} is still valid.
In practice, therefore, we propose to ignore those $\l$ under which $\df_\l>n$
and apply the LR criterion as usual.
A theoretically rigorous treatment is beyond the scope of the present paper,
which we intend to do in a future paper.
However, a systematic simulation study in the next section suggests that
the LR criterion still works surprisingly well and enjoys model selection consistency.  

\section{Numerical examples}\label{Secsimulation}
In this section, we present simulation studies for the LR criterion, 
compare the LR criterion to other methods, and also apply it to a real data set.
The regularization procedure we use is Lasso.
The Lasso solution paths are computed by the LARS algorithm of \cite{Efron:04}.
A widely-used method for choosing the Lasso parameter is GCV \citep{Craven:79,Tibshirani:96}
\beqn
\GCV_\l=\dfrac1n\dfrac{\|\v y-X\hat{\v \beta}_\l\|^2}{(1-\fr1n\DF_\l)^2}
\eeqn
where $\DF_\l:=\tr[X(X^\trp X+\l W^{-})^{-1}X^\trp\v y],\ W=\diag(|\hat\beta_j^\l|)$ 
and $W^{-}$ is a generalized inverse of $W$.
Another one is the BIC-type criterion of \cite{Wang:07} 
(although its variable selection consistency requires the oracle property,
a property not enjoyed by Lasso)
\beqn
\widetilde\BIC_\l=\log\dfrac{\|\v y-X\hat{\v \beta}_\l\|^2}{n}+\DF_\l\dfrac{\log n}{n}.
\eeqn
Note that $\hat{\v\beta}_\l\not=\hat{\v\beta}_{\S_\l}$.
The former is the Lasso estimator whereas the latter is the OLS estimator resulting from 
fitting model $\S_\l$ by least squares.
Our proposed criteria \eqref{equLR} and \eqref{BIC} are constructed based on $\hat{\v\beta}_{\S_\l}$, not $\hat{\v\beta}_\l$.
This is the essential difference between our approach and the others.

\paradot{Example 1: small $d$}
We consider the following example which is taken from \cite{Tibshirani:96}:
\beqn
y=\v x^\trp\v\beta+\sigma\epsilon
\eeqn
where $\v\beta=(3,\ 1.5,\ 0,\ 0,\ 2,\ 0,\ 0,\ 0)^\trp$,
$x_i$ are marginally $N(0,1)$ 
with the correlation between $x_i$ and $x_j$ equal to $0.5^{|i-j|}$,
$\epsilon\sim N(0,1)$.
We compare the performance of LR and BIC criterion to that of GCV and $\widetilde\BIC$.
The performance is measured by the frequency of underfitting, overfitting and correct fitting 
and average number of zero coefficients over 100 replications. 

Table \ref{table1} summarizes the simulation results for various factors $n$ and $\sigma$.
Although $\widetilde\BIC$ works slightly better than GCV, it still produces overfitted models most of the time.
BIC does a good job and LR outperforms the others.

\begin{table}[ht]
\caption{The small-$d$ case} 
\centering 
\begin{tabular}{cccccccc} 
\hline\hline 
$\sigma$   &   n  & Method 	  & Under-     	 	& \text{Correctly}  & \text{Overfitted(\%)}&Ave. No.\\
	   &      &               & fitted(\%) 	 	& \text{fitted(\%)} &                      &of zeros\\  \hline\hline
    1	   &  100 & \GCV	  &  0			&  0		    &	100		   & 1.57\\
	   &      &$\widetilde\BIC$&  0			&  3    	    &   97		   & 2.32\\
	   &      & \BIC	  &  0			&  89    	    &   11		   & 4.88\\
	   &      & \LR    	  &  0			&  97		    &    3		   & 4.97\\
\cline{2-7} 
	   &  200 & \GCV          &  0                	&  0                &   100                & 1.64\\
           &      &$\widetilde\BIC$&  0                	&  0                &   100                & 1.81\\
           &      & \BIC          &  0                	&  94               &   6                & 4.93\\
           &      & \LR           &  0                	&  100              &   0                  & 5\\
\hline\hline
   3      &   100  & \GCV          &  0                &  0                &   100                & 1.34\\
           &       &$\widetilde\BIC$&  0                &  0                &   100                & 1.53\\
           &       & \BIC          &  1                &  70                &   29                & 4.22\\
           &       & \LR           &  1                &  77               &   22                 & 4.37\\
\cline{2-7}
           &   200 & \GCV          &  0                &  0                &   100                & 1.69\\
           &       &$\widetilde\BIC$&  0                &  0                &   100                & 2.09\\
           &       & \BIC          &  0                &  91               &   9                  & 4.89\\
           &       & \LR           &  0                &  91               &   9                  & 4.90\\
[1ex] 
\hline
\end{tabular}
\label{table1} 
\end{table}

\paradot{Example 2: large $d$}
We consider cases of large $d$ in this example with $d=300$ and $n=100,\ 200,\ 500$.
We set up a {\it sparse recovery problem} in which most of coefficients are zero except $\b_{30}=\b_{60}=...=\b_{300}=10$.
The design matrix is simulated as in Example 1.
Table \ref{table2} summarizes the simulation results for various factors $n=100,\ 200,\ 500$ and $\sigma=1,\ 3$.
The LR criterion works surprisingly well in comparison with BIC and the others.

Let us take a closer look at the simulation results in Tables \ref{table1}-\ref{table2}.
Although the LR and BIC criteria are {\em asymptotically} equivalent to each other,
the finite-sample simulation study shows that the LR criterion works better than BIC.
A similar situation was also observed in \cite{Hutter:09} for subset selection.
This is probably because, 
contrarily to the BIC criterion,
the penalty term of the LR criterion is data-adaptive.
Some results in the model selection literature show that 
selection criteria with data-adaptive penalties are more encouraging than those with deterministic penalties; 
see \cite{Yang:05} and references therein.
We see that BIC seems to break down for the cases $d>n$
as it always produces overfitted models, but starts working well when $n>d$.
The $O_\P(1)$ term in \eqref{equlemma} plays an important role here:
it serves as a ``corrector" to BIC.
Note that BIC is just an approximation to the logarithm of posterior model probability \citep{Schwarz:78},
the approximation might be inaccurate if $n$ is not large enough relative to $d$.

\begin{table}[ht]
\caption{The large-$d$ case} 
\centering 
\begin{tabular}{cccccccc} 
\hline\hline 
$\sigma$   &   n  & Method 	  & Under-     	 	& \text{Correctly}  & \text{Overfitted(\%)}&Ave. No.\\
	   &      &               & fitted(\%) 	 	& \text{fitted(\%)} &                      &of zeros\\  \hline\hline
    1	   &  100 & \GCV	  &  0			&  0		    &	100		   & 90.20\\
	   &      &$\widetilde\BIC$& 0			&  0    	    &   100		   & 95.8\\
	   &      & \BIC    	  &  0			&  0		    &    100		   & 202.01\\
	   &      & \LR    	  &  0			&  30		    &    70		   & 288.24\\
\cline{2-7}
	   &  200 & \GCV          &  0                	&  0                &   100                & 87.51\\
           &      &$\widetilde\BIC$& 0                	&  0                &   100                & 89.45\\
	   &      & \BIC    	  &  0			&  0		    &    100		   & 102.02\\
           &      & \LR           &  0                	&  86               &   14                  & 289.83\\
\cline{2-7}
	   &  500 & \GCV          &  0                	&  0                &   100                & 97.51\\
           &      &$\widetilde\BIC$& 0                	&  0                &   100                & 104.45\\
	   &      & \BIC    	  &  0			&  40		    &    60		   & 287.30\\
           &      & \LR           &  0                	&  100               &   0                  & 290\\
\hline\hline
   3      &   100  & \GCV          &  0                &  0                &   100                & 78.35\\
           &       &$\widetilde\BIC$& 0                &  0                &   100                & 87.40\\
	   &       & \BIC    	   &  0		       &  0		   &   100		  & 202.04\\
           &       & \LR           &  0                &  18               &   82                 & 287.51\\
\cline{2-7}
           &   200 & \GCV          &  0                &  0                &   100                & 92.02\\
           &       &$\widetilde\BIC$& 0                &  0                &   100                & 96.51\\
	   &       & \BIC    	   &  0		       &  0		   &    100		   & 102.01\\
           &       & \LR           &  0                &  58               &   42                  & 289.29\\
\cline{2-7}
	   &  500 & \GCV          &  0                	&  0                &   100                & 93.31\\
           &      &$\widetilde\BIC$& 0                	&  0                &   100                & 96.52\\
	   &      & \BIC    	  &  0			&  35		    &    65		   & 288.35\\
           &      & \LR           &  0                	&  80               &   20                  & 289.75\\
[1ex] 
\hline
\end{tabular}
\label{table2} 
\end{table}

\paradot{Example 3: Prostate cancer data}
We consider a real data set in this example.
\cite{Stamey:89} studied the correlation between the level of
prostate antigen ({\it lpsa}) and a number of clinical measures in
men: log cancer volume ({\it lcavol}), log prostate weight ({\it lweight}), 
{\it age}, log of the amount of benign prostatic hyperplasia ({\it lbph}), 
seminal vesicle invasion ({\it svi}), log of capsular penetration ({\it lcp}), 
Gleason score ({\it gleason}), and percentage of Gleason scores 4 or 5 ({\it pgg45}). 
Following \cite{Tibshirani:96}, we assume a linear regression model between the response {\it lpsa} and the 8 covariates.
We want to select a parsimonious model for the sake of scientific insight into the response-covariate relationship.
 
\begin{figure*}
\centerline{\includegraphics[width=.6\textwidth]{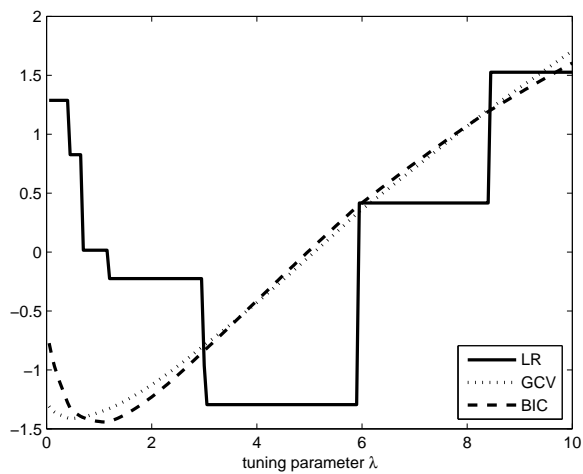}}
\caption{\label{figure2}
Prostate cancer data: $\LR_\l,\ \widetilde\BIC_\l$ and $\GCV_\l$.}
\end{figure*}

The data set of size 97 is standardized so that the intercept $\beta_0$ is excluded.
Figure \ref{figure2} presents the curves $\GCV_\l,\ \widetilde\BIC_\l,\ \LR_\l$
(1000 values of $\l$ ranging from 0.01 to 10 in increments of .01 were used to search for the optimal $\l$). 
The $\l$ selected by $\GCV,\ \widetilde\BIC$ are $.5$ and $1.1$, 
and the corresponding models are $\{1,\ 2,\ 3,\ 4,\ 5,\ 7,\ 8\}$, $\{1,\ 2,\ 3,\ 4,\ 5,\ 8\}$, respectively.
The LR criterion is minimized in the interval $(3.1,5.9)$.
Any value in this interval produces the same model $\S_\LR=\{1,\ 2,\ 5\}$.
The BIC of these models are $-19.20,\ -21.38,\ -25.19$, respectively.
That means the BIC also supports the choice of the LR criterion.
(Note however that this does not mean that the BIC is an optimal criterion).

\section{Conclusions and outlook}\label{secCon}
Regularization procedures are efficient methods for variable selection,
subject to a proper choice of shrinkage parameter.
By virtue of LoRP, a general-purpose principle for model selection,
the LR criterion for variable selection in linear regression analysis was proposed.
Variable selection consistency of the suggested criterion was pointed out
theoretically and experimentally.
Both theoretical and experimental results show that the proposed criterion
is a very encouraging procedure for variable selection problem, especially in high-dimensional settings.
Regularization procedures have now been extended to genaralized linear models and beyond,
we intend to extend our approach to such frameworks in future work.

\noindent---------------------

\noindent {\it M. N. Tran, Department of Statistics and Applied Probability, National University of Singapore, Singapore 117546.\\
E-mail: ngoctm@nus.edu.sg}

\section*{Appendix}\label{secAppendix}
\begin{proof}[Proof of Theorem \ref{LRconsistency}]
The main idea of the proof is taken from \cite{Chambaz:06}.
Let us denote by $\v z_i=(x_{i1},...,x_{id},y_i)$ the $i$-th observation and by
$\gamma(.,m,\sigma^2)$ the density of the Gaussian distribution with mean $m$ and variance $\sigma^2$.
Under model $\S$, the density of $\v z_i$ is $p_{\theta_\S}(\v z_i)=\gamma(y_i,\sum_{j\in\S}\beta_jx_{ij},\sigma^2)$.
The log-likelihood is
\beqn
l_n(\theta_\S)=\sum_{i=1}^n\log p_{\theta_\S}(\v z_i)=-\fr n2\log(2\pi)-\fr n2\log\sigma^2-\fr1{2\sigma^2}\sum_{i=1}^n(y_i-\sum_{j\in\S}\beta_jx_{ij})^2.
\eeqn
It is easy to see that
\beqn
\sup_{\theta\in\Theta(\S)}l_n(\theta)=-\fr n2\log\hat\sigma_\S^2-\fr n2(1+\log(2\pi)).
\eeqn
By \eqref{equlemma}, the loss rank of model $\S_\l$ now can be written as
\beqn
\LR_\l=-\sup_{\theta\in\Theta(\S_\l)}l_n(\theta)+\fr{\df_\l}{2}\log n + C(n) +O_\P(1)
\eeqn
where the constant term $C(n)=\fr n2\log n-\fr n2(1+\log(2\pi))$ is independent of $\S_\l$.

\paradot{No underestimation}
It is sufficient to prove that $\P(\S_{\hat\l_\LR}=\S)\to0$ for each $\S\not\supseteq \S_T$,
as there is only a finite number of such $\S$.
\bqa\label{above}
\P(\S_{\hat\l_\LR}=\S)&=&\P(\S_{\hat\l_\LR}=\S,\LR_{\hat\l_\LR}\leq\LR_{\l_n})\notag\\
&=&\P\Big(\fr1n\sup_{\theta\in\Theta(\S_{\hat\l_\LR})}l_n(\theta)-\fr1n\sup_{\theta\in\Theta(\S_{\l_n})}l_n(\theta)\geq\fr{\log n}{2n}(\df_{\hat\l_\LR}-\df_{\l_n})+o_\P(1),\ \S_{\hat\l_\LR}=\S\Big)\nonumber\\
&\leq&\P\Big(\fr1n\sup_{\theta\in\Theta(\S)}l_n(\theta)-\fr1n\sup_{\theta\in\Theta(\S_{\l_n})}l_n(\theta)\geq\fr{\log n}{2n}(|\S|-\df_{\l_n})+o_\P(1)\Big)\nonumber\\
&\leq& \P\Big(\fr1n\sup_{\theta\in\Theta(\S)}l_n(\theta)-\fr1n\sup_{\theta\in\Theta(\S_T)}l_n(\theta)\geq\fr{\log n}{2n}(|\S|-d^*)+o_\P(1)\Big)+P(\S_{\l_n}\not= \S_T)\nonumber\\
&\leq&\P\Big(\fr1n\sup_{\theta\in\Theta(\S)}l_n(\theta)-\fr1n l_n(\theta^*)\geq\fr{\log n}{2n}(|\S|-d^*)+o_\P(1)\Big)+P(\S_{\l_n}\not= \S_T)
\eqa
where $\theta^*\in\S_T$ denotes the true parameter. By the law of large numbers for
the supremum of the likelihood ratios (see, e.g., Lemma B1 of \cite{Chambaz:06}) 
\beqn
\fr1n\sup_{\theta\in\Theta(\S)}l_n(\theta)-\fr1n l_n(\theta^*)\to-\inf_{\theta\in\Theta(\S)}\KL(p_{\theta^*}\|p_\theta)\;\;\text{w.p.1.}
\eeqn
Because $\S\not\supseteq\S_T$, $\inf_{\theta\in\Theta(\S)}\KL(p_{\theta^*}\|p_\theta)>0$.
This, together with the fact that $\fr{\log n}{2n}(|\S|-d^*)\to0$ and Assumption (A1), 
shows that the left-hand side term of \eqref{above} goes to 0 as $n\to\infty$.

\paradot{No overestimation}
Fix an overfitted model $\S\supsetneq\S_T$, let us denote by
\beqn
H(\theta):=\KL(p_{\theta^*}\|p_\theta)=E[\fr1n(l_n(\theta^*)-l_n(\theta))]\geq0\;\forall\theta\in\Theta(\S)
\eeqn
($H(\theta)$ is not necessarily positive) and $h_n(\theta):=\fr{l_n(\theta)-l_n(\theta^*)}{H(\theta)^{1/2}}$ with convention $\fr00=0$. 
For every $\theta\in\Theta(\S)$
\bqa
l_n(\theta)-l_n(\theta^*)+nH(\theta)&=&l_n(\theta)-l_n(\theta^*)-E[l_n(\theta)-l_n(\theta^*)]\nonumber\\
&=&H(\theta)^{1/2}(h_n(\theta)-Eh_n(\theta))\nonumber\\
&\leq&H(\theta)^{1/2}\sup_{\nu\in\Theta(\S)}(h_n(\nu)-Eh_n(\nu)).\label{above1}
\eqa
By $\Theta(\S_T)\subset\Theta(\S)$ and the property of supremum, for every $\epsilon>0$ 
there exists $\theta_0\in\Theta(\S)$ such that
\beq\label{above2}
\sup_{\theta\in\Theta(\S)}(l_n(\theta)-l_n(\theta^*))\leq l_n(\theta_0)-l_n(\theta^*)+\epsilon
\eeq
and also
\beq\label{above3}
l_n(\theta_0)-l_n(\theta^*)\geq0.
\eeq
From \eqref{above2} and \eqref{above1}
\beq\label{above4}
\sup_{\theta\in\Theta(\S)}(l_n(\theta)-l_n(\theta^*))\leq H(\theta_0)^{1/2}\sup_{\theta\in\Theta(\S)}(h_n(\theta)-Eh_n(\theta))+\epsilon.
\eeq
From \eqref{above3} and \eqref{above1}
\beqn
nH(\theta_0)\leq l_n(\theta_0)-l_n(\theta^*)+nH(\theta_0)\leq H(\theta_0)^{1/2}\sup_{\theta\in\Theta(\S)}(h_n(\theta)-Eh_n(\theta))
\eeqn
or
\beq\label{above5}
nH(\theta_0)^{1/2}\leq \sup_{\theta\in\Theta(\S)}(h_n(\theta)-Eh_n(\theta)).
\eeq
Now, since $\epsilon>0$ was chosen arbitrarily, \eqref{above4} and \eqref{above5} yield
\beq\label{result}
\sup_{\Theta(\S)}l_n(\theta)-\sup_{\Theta(\S_T)}l_n(\theta)\leq\sup_{\Theta(\S)}\{l_n(\theta)-l_n(\theta^*)\}\leq\fr1n\left(\sup_{\theta\in\Theta(\S)}(h_n(\theta)-Eh_n(\theta))\right)^2.
\eeq 
We need the following bounded law of the iterated logarithm which is a consequence of Theorem 4.1, \cite{Dudley:83} or Lemma B2, \cite{Chambaz:06}.
\begin{lemma}\label{lemma5}
There is a finite constant $C$ so that
\beqn
\limsup_n\dfrac{\sup_{\theta\in\Theta(\S)}|h_n(\theta)-Eh_n(\theta)|}{\sqrt{n\log\log n}}\leq C\;\;\text{w.p.1}.
\eeqn
\end{lemma}
Now for every overfitted model $\S\supsetneq\S_T$, it is sufficient to prove that $\P(\S_{\hat\l_\LR}=\S)\to0$.
In fact,
\bqa\label{cucku}
&&\P(\S_{\hat\l_\LR}=\S)=\P(\S_{\hat\l_\LR}=\S,\ \LR_{\hat\l_\LR}\leq\LR_{\l_n})\notag\\
&\leq&\P\left(\sup_{\Theta(\S)}l_n(\theta)-\sup_{\Theta(\S_{\l_n})}l_n(\theta)\geq\fr{\log n}{2}(|\S|-\df_{\l_n})+O_\P(1)\right)\notag\\
&\leq&\P\left(\sup_{\Theta(\S)}l_n(\theta)-\sup_{\Theta(\S_T)}l_n(\theta)\geq\fr{\log n}{2}(|\S|-d^*)+O_\P(1)\right)+\P(\S_{\l_n}\not=\S_T)\notag\\
&=&\P\left(\Big[\dfrac{\log\log n}{\fr{d^*}{2}\log n}\Big]\Big[\dfrac{\sup_{\Theta(\S)}l_n(\theta)-\sup_{\Theta(\S_T)}l_n(\theta)}{\log\log n}\Big]\geq\fr{|\S|}{d^*}-1+o_\P(1)\right)+\P(\S_{\l_n}\not=\S_T)\notag\\
&\leq&\P\left(\Big[\dfrac{\log\log n}{\fr{d^*}{2}\log n}\Big]\Big[\dfrac{\sup_{\Theta(\S)}|h_n(\theta)-Eh_n(\theta)|}{\sqrt{n\log\log n}}\Big]^2\geq\fr{|
\S|}{d^*}-1+o_\P(1)\right)+\P(\S_{\l_n}\not=\S_T)
\eqa 
where the last inequality follows from \eqref{result}.
Observe that $|\S|>d^*$ as $\S\supsetneq\S_T$.
This, together with Lemma \ref{lemma5} and the fact that ${\log\log n}/(\fr{d^*}{2}\log n)\to0$, implies that
the first probability of \eqref{cucku} goes to zero.
The second probability of \eqref{cucku} also goes to zero because of Assumption (A1).
This completes the proof. 
\end{proof}


\bibliographystyle{apalike}

\begin{thebibliography}{}
\bibitem[Akaike(1973)]{Akaike:73}
Akaike H. (1973). 
\newblock Information theory and an extension of the maximum likelihood principle.
\newblock In {\em Proc. 2nd International Symposium on Information Theory},
  pages 267--281, Budapest, Hungary, Akademiai Kaid\'o.

\bibitem[Bartlett et~al.(2002)Bartlett, Boucheron, and Lugosi]{Bartlett:02}
Bartlett P., Boucheron S. and Lugosi G. (2002). 
\newblock Model selection and error estimation.
\newblock {\em Machine Learning}, 48, 85--113.

\bibitem[Burnham and Anderson(2002)]{Burnham:02}
Burnham K. P. and Anderson D. (2002).
\newblock {\em Model selection and multimodel inference: a practical
  information-theoretic approach}.
\newblock New York, Springer.

\bibitem[Candes and Tao(2007)]{Candes:07}
Candes E. and Tao T. (2007).
\newblock The dantzig selector: statistical estimation when $p$ is much larger
  than $n$ (with discussion).
\newblock {\em The Annals of Statistics}, 35, 2313--2351.

\bibitem[Chambaz(2006)]{Chambaz:06}
Chambaz A. (2006).
\newblock Testing the order of a model.
\newblock {\em The Annals of Statistics}, 34, 1166--1203.

\bibitem[Craven and Wahba(1979)]{Craven:79}
Craven P. and Wahba G. (1979).
\newblock Smoothing noisy data with spline functions: estimating the correct
  degree of smoothing by the methods of generalized cross-validation.
\newblock {\em Numerische Mathematik}, 31, 377--403.

\bibitem[Dudley and W.Philipp(1983)]{Dudley:83}
Dudley R. M. and Philipp W. (1983).
\newblock Invariance principles for sums of banach space valued random elements
  and empirical processes.
\newblock {\em Z. Wahrsch. Verw. Gebiete}, 62, 509--552.

\bibitem[Efron et~al.(2004)Efron, Hastie, Johnstone, and Tibshirani]{Efron:04}
Efron B., Hastie T., Johnstone I. and Tibshirani R. (2004).
\newblock Least angle regression.
\newblock {\em The Annals of Statistics}, 32, 407--499.

\bibitem[Fan and Li(2001)]{Fan:01}
Fan J. and Li R. (2001).
\newblock Variable selection via nonconcave penalized likelihood and its oracle
  properties.
\newblock {\em Journal of the American Statistical Association}, 96, 1348--1360.

\bibitem[Friedman(2008)]{Friedman:08}
Friedman J. H. (2008).
\newblock Fast sparse regression and classification.
\newblock URL {http://www-stat.stanford.edu/~jhf/ftp/GPSpaper.pdf}.

\bibitem[Hutter(2007)]{Hutter:07}
Hutter M. (2007).
\newblock The loss rank principle for model selection.
\newblock In {\em Proc. 20th Annual Conf. on Learning Theory ({COLT'07})},
  volume 4539 of {\em LNAI}, 589--603, San Diego, Springer, Berlin.

\bibitem[Hutter and Tran(2010)]{Hutter:09}
Hutter M. and Tran M. N. (2010).
\newblock Model selection with the loss rank principle.
\newblock {\em Computational Statistics and Data Analysis}, 54,
  1288--1306.

\bibitem[Koltchinskii(2001)]{Koltchinskii:01}
Koltchinskii V. (2001).
\newblock Rademacher penalties and structural risk minimization.
\newblock {\em IEEE Transaction on Information Theory}, 47,
  1902--1914.

\bibitem[Leng et~al.(2006)Leng, Lin, and Wahba]{Leng:06}
Leng C., Lin Y. and Wahba G. (2006).
\newblock A note on the lasso and related procedures in model selection.
\newblock {\em Statistica Sinica}, 16, 1273--1284.

\bibitem[Meinshausen and Buhlmann(2006)]{Meinshausen:06}
Meinshausen N. and Buhlmann P. (2006).
\newblock Consistent neighbourhood selection for high-dimensional graphs the
  lasso.
\newblock {\em The Annals of Statistics}, 34, 1436--1462.

\bibitem[Miller(1990)]{Miller:90}
Miller A. (1990).
\newblock {\em Subset Selection in Regression}.
\newblock Chapman \& Hall/CRC.

\bibitem[Poetscher and Leeb(2009)]{Poetscher:09}
Poetscher B. M. and Leeb H. (2009).
\newblock On the distribution of penalized maximum likelihood estimators: The
  lasso, scad, and thresholding.
\newblock {\em Journal of Multivariate Analysis}, 100, 2065--2082.

\bibitem[Schwarz(1978)]{Schwarz:78}
Schwarz G. (1978).
\newblock Estimating the dimension of a model.
\newblock {\em The Annals of Statistics}, 6, 461--464.

\bibitem[Shao(1997)]{Shao:97}
Shao J. (1997).
\newblock An asymptotic theory for linear model selection.
\newblock {\em Statistica Sinica}, 7, 221--264.

\bibitem[Stamey et~al.(1989)Stamey, Kabalin, McNeal, Johnstone, Freiha,
  Redwine, , and Yang]{Stamey:89}
Stamey T., Kabalin J., McNeal J., Johnstone I., Freiha F., Redwine E. and Yang N. (1989).
\newblock Prostate specific antigen in the diagnosis and treatment of
  adenocarcinoma of the prostate ii. radical prostatectomy treated patients.
\newblock {\em Journal of Urology}, 16, 1076--1083.

\bibitem[Tibshirani(1996)]{Tibshirani:96}
Tibshirani R. (1996).
\newblock Regression shrinkage and selection via the lasso.
\newblock {\em Journal of the Royal Statistical Society. Series B}, 58, 267--288.

\bibitem[Tran(2009)]{Tran:09a}
Tran M. N. (2009). 
\newblock Penalized maximum likelihood principle for choosing ridge parameter.
\newblock {\em Communication in Statistics: Simulation and Computation}, 38, 1610--1624.

\bibitem[Tran and Hutter(2010)]{Tran:09b}
Tran M. N. and Hutter H. (2010).
\newblock Model selection by loss rank for discrete data.
\newblock {\em Submitted}.

\bibitem[Wang et~al.(2007)Wang, Li, and Tsai]{Wang:07}
Wang H., Li R. and Tsai C. L. (2007).
\newblock Tuning parameter selectors for the smoothly clipped absolute
  deviation method.
\newblock {\em Biometrika}, 94, 553--568.

\bibitem[Yang(2005)]{Yang:05}
Yang Y. (2005).
\newblock Can the strengths of AIC and BIC be shared? A conflict between model indentification and regression
estimation
\newblock {\em Biometrika}, 92, 937--950.

\bibitem[Zhao and Yu(2006)]{Zhao:06}
Zhao P. and Yu B. (2006).
\newblock On model selection consistency of lasso.
\newblock {\em Journal of Machine Learning Research}, 7, 2541--2563.

\bibitem[Zou(2006)]{Zou:06}
Zou H. (2006).
\newblock The adaptive lasso and its oracle properties.
\newblock {\em Journal of the American Statistical Association}, 101, 1418--1429.

\bibitem[Zou et~al.(2007)Zou, Hastie, and Tibshirani]{Hui:07}
Zou H., Hastie T. and Tibshirani R. (2007).
\newblock On the degrees of freedom of the lasso.
\newblock {\em The Annals of Statitics}, 35, 2173--2192.

\end{thebibliography}

\end{document}